\documentclass[usenatbib]{mn2e}
\usepackage{natbib}
\bibpunct{(}{)}{;}{a}{}{,} 
\usepackage{xspace}
\usepackage{amssymb}
\usepackage{amsmath}
\usepackage{graphicx}

\begin{document}
\title[Interferometer predictions with triangulated images]{Interferometer 
       predictions with triangulated images: solving the multi-scale problem}

\author[C.~Brinch \& C.~P.~Dullemond]{C.~Brinch$^1$
\thanks{E-mail: brinch@nbi.dk} and C.~P.~Dullemond$^{2,3}$ \\
$^1$Centre for Star and Planet Formation (Starplan) and Niels Bohr Institute \\
University of Copenhagen, Juliane Maries Vej 30, 2100 Copenhagen \O, Denmark\\
$^2$Institut f\"ur Theoretische Astrophysik Zentrum f\"ur Astronomie der 
Universit\"at Heidelberg\\ Albert-Ueberle-Str. 2, 69120 Heidelberg, Germany\\
$^3$Max-Planck-Institute f\"ur Astronomie K\"onigstuhl 17,69117 Heidelberg, 
Germany
}

\maketitle

\begin{abstract}
	Interferometers play an increasingly important role for spatially resolved 
	observations. If employed at full potential, interferometry can probe an 
	enormous dynamic range in spatial scale. Interpretation of the observed 
	visibilities requires the numerical computation of Fourier integrals over 
	the synthetic model images. To get the correct values of these integrals, 
	the model images must have the right size and resolution. Insufficient care 
	in these choices can lead to wrong results. We present a new general-purpose 
	scheme for the computation of visibilities of radiative transfer images. Our 
	method requires a model image that is a list of intensities at arbitrarily 
	placed positions on the image-plane. It creates a triangulated grid from these 
	vertices, and assumes that the intensity inside each triangle of the grid is 
	a linear function. The Fourier integral over each triangle is then evaluated 
	with an analytic expression and the complex visibility of the entire image is 
	then the sum of all triangles. The result is a robust Fourier transform that 
	does not suffer from aliasing effects due to grid regularities.  The method 
	automatically ensures that all structure contained in the model gets reflected 
	in the Fourier transform.
\end{abstract}

\begin{keywords}
Techniques: image processing, Techniques: interferometric
\end{keywords}

\section{Introduction}
The technique of interferometry has a long history in radio astronomy and gains 
more and more popularity also at other wavelengths. In the millimetre and 
sub-millimetre domain arrays such as the SMA, Plateau de Bure and CARMA allow, 
for instance, young stellar objects and protoplanetary disks to be spatially 
resolved down to a few tens of AU. And soon, ALMA will achieve few-AU resolution 
at wavelengths ranging from 0.3 to 3 mm. In the mid- and near-infrared optical 
interferometry is maturing as well and has provided new insights into the 
physics of protoplanetary disks and active galactic nuclei. The interpretation 
of these data, however, often requires detailed comparisons with theoretical 
models. Typically a radiative transfer model is produced of the object of 
interest, and the results compared to the observations. This paper is about this 
process of comparing models to interferometric measurements.

Interferometers probe the image of the object on the sky in the Fourier plane. 
Rather than measuring pixel-by-pixel intensities and thus immediately yielding 
an image for the observer to interpret, in radio and millimeter interferometry each pair of telescopes measures the so-called 'complex visibility' (the normalized correlation function between the signals measured by the two telescopes). In optical and infrared interferometers usually only the amplitude (not the phase) of the visibility is measured, which is in fact the ratio of the correlated flux density to the total flux density. According to the van Cittert-Zernike theorem the complex visibility as a function of baseline coordinates (u,v) is equal to the Fourier transform of the image on the sky divided by its total flux density. For each combination of three 
telescopes one can measure a ``closure phase'' which also directly follows from 
the complex Fourier values belonging to each telescope pair. Interferometry 
measurements are thus measurements in Fourier space, usually called the 
uv-plane. If sufficient baselines are available, i.e., the uv-plane is 
sufficiently well covered, then the inverse Fourier transform can be carried out 
and an image reconstructed. However, often the uv-plane is sparsely covered and 
image reconstruction is non-unique. In such cases, any model comparison will have to 
take place in the uv-plane itself, and model images must be Fourier transformed 
to the uv-plane before comparison can take place. Also for the case of a high
uv-coverage, this ``forward method'' (adapting the model to the observations) 
can also be useful for predicting the feasibility of observing particular 
objects and phenomena.

In some cases where the astronomical source has a simple structure which
can adequately be described by an algebraic expression (e.g., point, sphere, disk,
cylinder, ring, etc.), the complete Fourier transform is easily calculated 
analytically and used to model the data \citep{1999ASPC..180..335P,2014arXiv1401.4984M}.
However, if a numerical model (typically the output from a radiative transfer code)
is used to describe the source, the Fourier transform needs to be calculated numerically. The task of numerically calculating a Fourier transform of a model image may 
seem trivial. Algorithms such as Fast Fourier Transform (FFT) can do this with
high precision and speed. It turns out, however, that for models that involve 
a large dynamic range in spatial scale this task can be difficult. For 
example, the problem of a collapsing molecular cloud core of $10^4$ AU size,
with a proto-stellar disk inside of 100 AU size which surrounds a protostar of 
0.1 AU size covers already 5 orders of magnitude in spatial scale. Although 
current interferometers are not able to observe all of these scales simultanously, 
it is still possible to cover 2-3 orders of magnitude in spatial scale with ALMA.
Observations of a molecular line and the dust continuum will record large scale 
emission in the line centre and small scale emission in the line wings and 
surrounding continuum. Calculating the uv-plane image is then not trivial at all, 
and doing so without great care will inevitably lead to errors. For example, one would need to use sufficient padding with blank space around the source model in order to avoid mirror images in the Fourier transform. In this paper we 
will describe a new method of computing synthetic uv-plane ``observations'' which 
are extremely robust and yield proper results without much care. The method we 
present can easily be implemented into existing radiative transfer codes or it 
can be made into a stand-alone subroutine that can post-process the output from 
ray-tracing codes. All examples of the method presented in this paper has been 
made using a customised parallel version of the public available LIME code 
\citep{2010A&A...523A..25B}.

\section{Unstructured images}\label{grid}
\begin{figure*}
\centerline{\includegraphics[width=16cm]{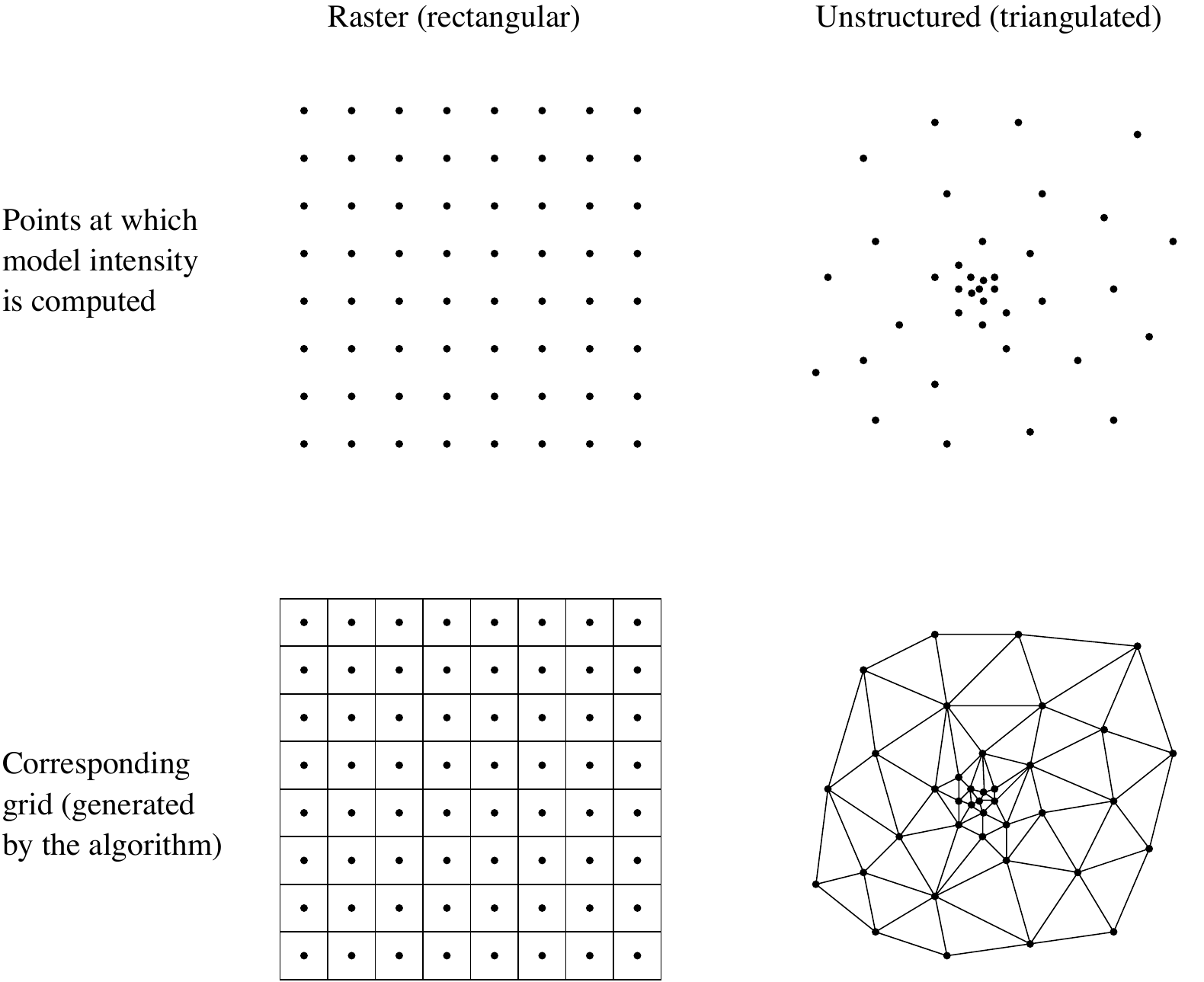}}
\caption{\label{fig-imagegrids} Schematic comparison of standard image pixel
   arrangement (left) and our unstructured pixel arrangement (right). The dots
   are the locations where the intensity $I_\nu$ is given (upper panels). For
   the standard pixel arrangements these correspond to the average values of
   square pixels, i.e. the dots are located in the middle of their pixel (lower
   left). For our unstructured pixel arrangements the dots are the corners of
   triangular ``pixels''. The intensity $I_\nu$ is linearly interpolated inside
   each triangle. The user of our algorithm only needs to provide the 
   coordinates in the image plane of a set of points and their corresponding 
   $I_\nu$ values (upper right). The algorithm will then produce the 
   corresponding triangulation (lower right) automatically.}
\end{figure*}

The main idea of our method is to provide model images not as ``raster images'' 
(Fig.~\ref{fig-imagegrids}-left) but as ``unstructured grid images'' 
(Fig.~\ref{fig-imagegrids}-right). 

Normally when one makes model images one computes the intensity on a regular
rectangular grid (a raster), like a normal digital photo. One thus obtains a
list of intensities as a function of $x-$ and $y-$ integer positions: $I_{i,j}$ 
with $1\le i\le N_x$ and $1\le j\le N_y$ where $N_xN_y$ is the total number of 
pixels. Each such intensity belongs to location $(x_i,y_i)$ on the image where 
$x_i=x_0+i\Delta x$ and $y_j=y_0+j\Delta y$, where $\Delta x$ and $\Delta y$ 
determine the spatial resolution. Usually $\Delta x=\Delta y$.  The largest 
spatial scale that can be sampled with such images is $N_x\Delta x$ in 
x-direction and $N_y\Delta y$ in y-direction. The smallest spatial scale is 
$\Delta x$ in x-direction and $\Delta y$ in y-direction. When making such an 
image for a model that covers a large dynamic range in spatial scale one must 
choose $\Delta x$, $\Delta y$ small enough and $N_x$ and $N_y$ large enough to 
encompass all these scales. In fact, when computing a meaningful Fourier 
transform for interferometry one must make sure that $N_x\Delta x$ and 
$N_y\Delta y$ are {\em substantially larger} than the largest scales you can 
pick up with your interferometer. This is because FFT assumes cyclic symmetry in 
$x$ and $y$ (leading to mirror copies of your image at regular intervals) 
whereas in reality this is not true. A proper Fourier transform thus always 
requires large enough $N_x$ and $N_y$, even for a single baseline. If one wishes 
to make {\em one} image for multiple baselines then this puts even stronger 
lower limits to $N_x$ and $N_y$.  In any event, one always has to take extreme 
care to choose $\Delta x$, $\Delta y$, $N_x$ and $N_y$ properly.

The idea we propose here is to produce images on an ``unstructured grid'' in the 
image plane (Fig.~\ref{fig-imagegrids}-right). This sounds much more complex 
than it actually is. The user of our method must generate a set of points 
$(x_i,y_i)$ on the image plane (Fig.~\ref{fig-imagegrids}-upper-right), compute 
the intensity $I_{\nu,i}$ corresponding to each of these points, and provide the 
set $(x_i,y_i,I_{\nu,i})$ for $1\le i\le N$, as well as a set of uv-spacings
corresponding to the telescope baselines of interest. Our algorithm will then 
generate a proper triangulation (Fig.~\ref{fig-imagegrids}-lower-right), and 
compute the complex visibility values for each baseline, as we shall describe in 
the next section. The only thing the user must take care of is to assure that 
the set of image grid points $(x_i,y_i)$ properly map the entire model. Regions 
where there are small-scale structures require a denser sampling than regions of 
large scale smooth structures. If we take the example of a collapsing molecular 
cloud core with a central star+disk again one should assure that there are 
sufficient points probing the very small (100 AU) disk and an equally large 
number of points covering the large (10000 AU) scale infalling envelope. If one 
randomly places points with a probability distribution such that there are 
statistically as much points between 1 and 100 AU as there are between 100 and 
10000 AU, then one has presumably a proper sampling. An example of such a 
probability distribution is

\begin{equation}
P(r)dr \propto \frac{dr}{r}
\end{equation}
normalised such that 
\begin{equation}
\int_{r_{\mathrm{in}}}^{r_{\mathrm{out}}}P(r)dr = 1.
\end{equation}

\begin{figure}
\centerline{\includegraphics[width=6cm]{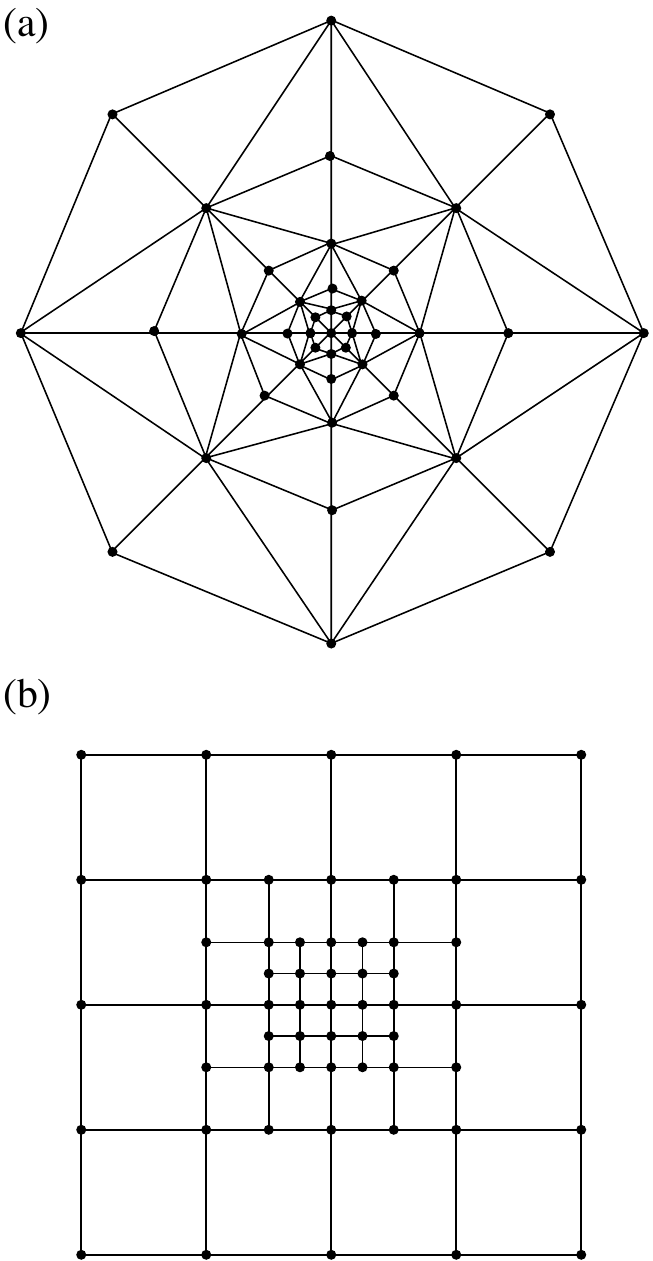}}
\caption{\label{fig-altgrids} With the general formalism of unstructured grids 
   (i.e.~specifying a set $(x_i,y_i,I_{\nu,i})$ for $1\le i\le N$) one can also
   produce regular grids that still resolve spatial scales properly. For
   centrally condensed systems such as protoplanetary/proto-stellar disks or
   collapsing clouds onto single stars one could arrange the pixels in circles
   (upper). For more general clumpy media one could use a patch-style grid
   (lower). The danger with these grid is, however, that the regularity could
   lead to dangerous ``aliasing'' effects, in particular with the patch-style
   gridding. A semi-random ordering of points as in
   Fig.~\ref{fig-imagegrids}-right is always safer and is preferred if 
   possible.}
\end{figure}

In principle one can also make more regular arrangements of pixels with this 
scheme which also resolve all spatial scales properly, see 
e.g.~Fig.~\ref{fig-altgrids}.

The optimal solution is to have a line of sight pass through each model grid 
cell. For a regular $N_xN_y$ computational grid (in 2D) this would result in the
traditional raster image. However, for an AMR style refined grid, the resulting
intensity point distribution would be similar to that of 
Fig.~\ref{fig-altgrids}. In 3D codes which already make use of a randomly 
sampled grid, e.g.,\ LIME, one would simply trace a line of sight for each grid 
point position projected onto the image plane. This automatically ensure that 
all structure of the model is probed by the ray-tracer.

\section{Computing the Fourier transform}\label{sec:FT}
Once a set of intensities $I_{\nu,i}$, whether structured or unstructured, has 
been obtained, our algorithm will calculate the visibilities. 

\begin{figure}
\centerline{\includegraphics[width=8.5cm]{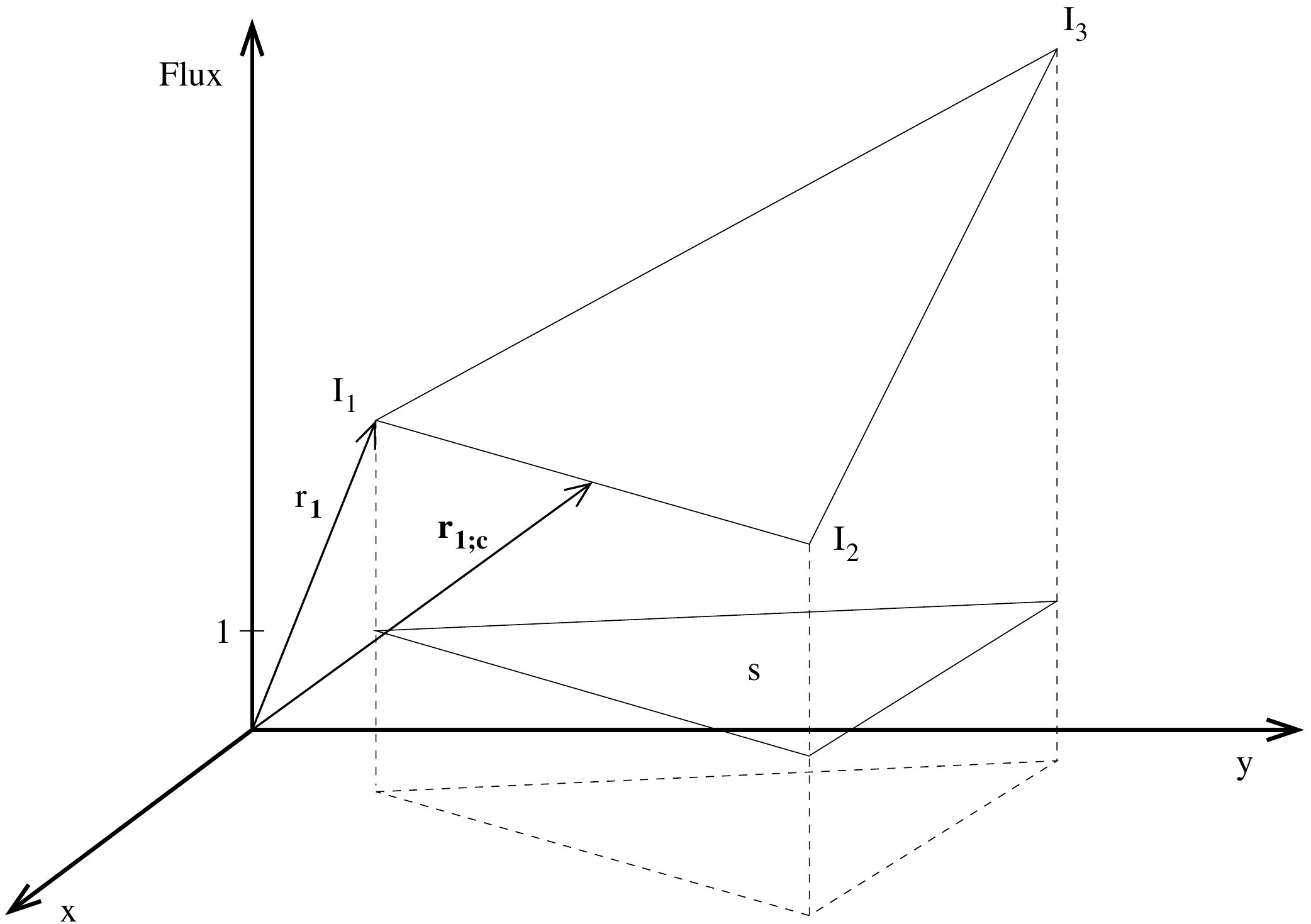}}
\caption{The geometry of a \emph{trixel} as used throughout Sec.~\ref{sec:FT}.}
\label{int_trian}
\end{figure}

The first step of the algorithm calculates the Delaunay triangulation of the
point set ($x_i,y_i$). The Delaunay triangulation is defined as the 
triangulation in which none of the points ($x_i,y_i$) lies inside any of the 
triangle circumcircles. The resulting triangulation has an intensity measurement 
at each triangle vertex. These triangles are the equivalent to the pixels in a 
raster map and we will refer to them as triangular pixels or \emph{trixels}. An 
example of the Delaunay triangulation is shown in the lower right panel of
Fig.~\ref{fig-imagegrids}. Algorithms for constructing the Delaunay 
triangulation are readily available in many scientific computation packages, and 
we use the Matplotlib\footnote{\emph{http://matplotlib.org}} Delaunay procedure. 
Any other similar tool, such as IDLs TRIANGULATE or the command line tool QHULL
\citep{Barber1996} can be used as well. One can also write custom triangulation 
code, for instance based on the method by~\citet{Lee1980}. An extensive 
discussion on how to implement Delaunay triangulations can be found in 
\citet{Springel2010}. Throughout the remainder of this section we refer the 
reader to Fig.~\ref{int_trian}, showing the geometry of a \emph{trixel}.

The two-dimensional Fourier transformation of an intensity distribution $I(x,y)$
is given by

\begin{eqnarray}\label{fourier}
F(\mathbf{w})=\iint^\infty_{-\infty} I(\mathbf{r}) e^{-i(\mathbf{w\cdot r})}d\mathbf{r},
\end{eqnarray}

where $\mathbf{w}=u\hat{\mathbf{x}}+v\hat{\mathbf{y}}$ and $\mathbf{r}=x 
\hat{\mathbf{x}}+y\hat{\mathbf{y}}$. We can split the Fourier transform into a 
sum of Fourier transforms  of the individual triangles because of the linearity 
of the transformation. Thus the task is reduced to calculating the Fourier 
transform of a single three sided polygon. If we at first make the simplifying 
assumption that I(x,y) is constant over the face of the triangle, e.g., by using 
the (weighted) mean of the three intensity values of the triangle vertices, we 
can follow the derivation in~\citet{1991ITAP...39..252H} to obtain the solution

\begin{eqnarray}\label{houshmand}
F_\Delta(\mathbf{w}) = \sum_{n=1}^3 I_\Delta 
	\frac{\hat{\mathbf{z}}\times \mathbf{l}_n \cdot \mathbf{l}_{n-1}}
	     {(\mathbf{w}\cdot\mathbf{l}_n)(\mathbf{w}\cdot\mathbf{l}_{n-1})}
	e^{-i(\mathbf{w}\cdot\mathbf{r}_n)}      
\end{eqnarray}

where $\Delta$ denotes a single triangle with vertex coordinate vectors 
$\mathbf{r}_n$ and $\mathbf{l}_n=\mathbf{r}_{n+1}-\mathbf{r}_n$ is the vector 
along the $n$'th edge. The vertices and edges are enumerated counter clockwise. 
We then obtain the Fourier transform of the entire image by summing over all 
triangles,

\begin{eqnarray}\label{sum_tri}
F(\mathbf{w})=\sum_{all\ \Delta} F_\Delta(\mathbf{w}).
\end{eqnarray}

This formula is simple to work with and gives a good representation of the image 
in uv-space, but it does assume a single averaged value for each triangle. If 
the underlying model gets properly sampled by a ray-tracer, the intensity can be 
assumed to vary linearly between the vertices of the triangles and therefore we 
can assume that the face of a triangle is described by a plane spanned by the 
intensity at each of the three vertices. Now in order to introduce a linear 
variation of the intensity over the face of a triangle, we note that the 
following distribution over the triangular patch describes the plane spanned by 
the intensity values at the vertices,

\begin{eqnarray}\label{linvar}
I(\mathbf{r})=\sum_{n=1}^{3} I_n \frac{\hat{\mathbf{z}}\times\mathbf{l}_{n+1}
	\cdot (\mathbf{r}-\mathbf{r}_{n+1})}{2A} s(\mathbf{r}),
\end{eqnarray}

where $A$ is the area and $s$ is the shape function of the triangle,

\begin{eqnarray}
s(\mathbf{r}) = \left \{ \begin{array}{l l}
	1, \qquad & \mathbf{r}\in \Delta \\
	0, \qquad & \mathrm{otherwise}
	\end{array} \right .
\end{eqnarray}

By inserting Eq.~\ref{linvar} into Eq.~\ref{fourier}, we get the expression for 
the Fourier transform of the triangular patch

\begin{eqnarray}\label{mcinturff1}
F(\mathbf{w})= \sum_{n=1}^3 I_n \frac{\hat{\mathbf{z}}\times\mathbf{l}_{n+1}}{2A} \cdot
	\iint_\Delta (\mathbf{r}-\mathbf{r}_{n+1}) e^{-i(\mathbf{w}\cdot\mathbf{r}_n)} dA,
\end{eqnarray}

where the shape function has been eliminated by letting the domain of the 
integral be the support of triangular patch. The solution to the integral part 
of Eq.~\ref{mcinturff1} is derived in~\citet{McInturff:1991de} and we will 
proceed to quote the resulting expression

\begin{equation}\label{mcinturff2}
\begin{split}
 \iint_\Delta & (\mathbf{r}-\mathbf{r}_{n+1}) e^{-i(\mathbf{w}\cdot\mathbf{r}_n)} dA  = \\ 
&\frac{1}{|\mathbf{w}|^2} \sum_{m=1}^3 e^{-i(\mathbf{w}\cdot\mathbf{r}_{m;c})}  \\	
& \cdot \left \{ \left [ \hat{\mathbf{z}}\times\mathbf{l}_{m} + \left (i\mathbf{r}_{m;c}-i\mathbf{r}_{n+1} - \frac{2\mathbf{w}}{|\mathbf{w}|^2} \right ) \right .  \hat{\mathbf{z}}\cdot\mathbf{l}_m\times\mathbf{w} \right ] \\
& \left . j_0 \left (\frac{\mathbf{w}\cdot\mathbf{l}_m}{2}\right )-\mathbf{l}_m\frac{\hat{\mathbf{z}}\cdot\mathbf{l}_m\times\mathbf{w}}{2}j_1\left (\frac{\mathbf{w}\cdot\mathbf{l}_m}{2} \right ) \right \}
\end{split}
\end{equation}

$\mathbf{r}_{m;c}$ denotes the centre point of the $m$'th edge and $j_0$ and 
$j_1$ are the Bessel functions of the first kind of order $0$ and $1$, 
respectively. Again we obtain the Fourier transform of the entire image by 
summing over all individual triangles. Equation~\ref{mcinturff1} gives us a 
\emph{complete} set of visibilities from which we can select a subset that 
corresponds to a set of observed $u,v$-points or we can predict the 
interferometer response at different antenna configurations and integration 
times. This Fourier transform does not suffer from aliasing due to the 
regularity of the pixels since the triangles are randomly oriented (unless, of 
course, the triangle vertices are sampled at regular intervals) and we do not 
need a taper region to avoid mirror copies of the image which is needed when 
using FFT because FFT assumes periodic boundary conditions. The resulting 
Fourier components can easily be stored in the standard uv-FITS format 
\citep{2001A&A...376..359H}. The drawback of the Fourier 
transform~(\ref{mcinturff1}) is that it is not fast to compute. The method 
presented here takes O(N$\times$M) operations, where $M\leq N$, while FFT can be 
carried out in O(N $\log$ N) operations.

\begin{figure*}
\centerline{\includegraphics[width=18cm]{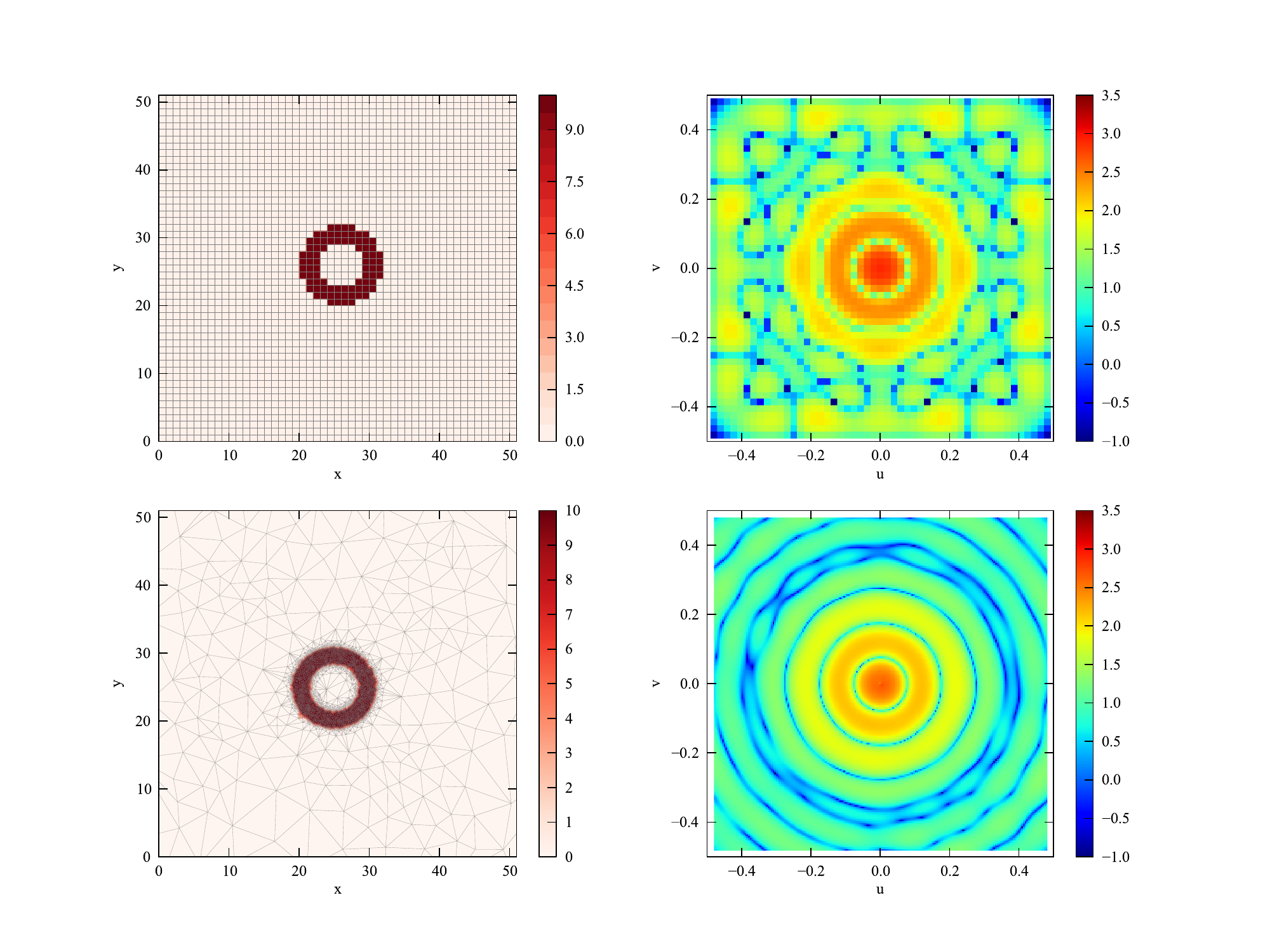}} 
\caption{The figure shows an image of an annulus (left) and the corresponding 
   Fourier amplitudes (right). In the top row, a normal FFT algorithm has been 
   applied to a regular pixel image, whereas in the bottom row, the method 
   presented in this paper has been applied to an irregular triangulated image.}
   \label{transforms1}
\end{figure*}

\section{Comparison of methods}
\begin{figure*}
\centerline{\includegraphics[width=18cm]{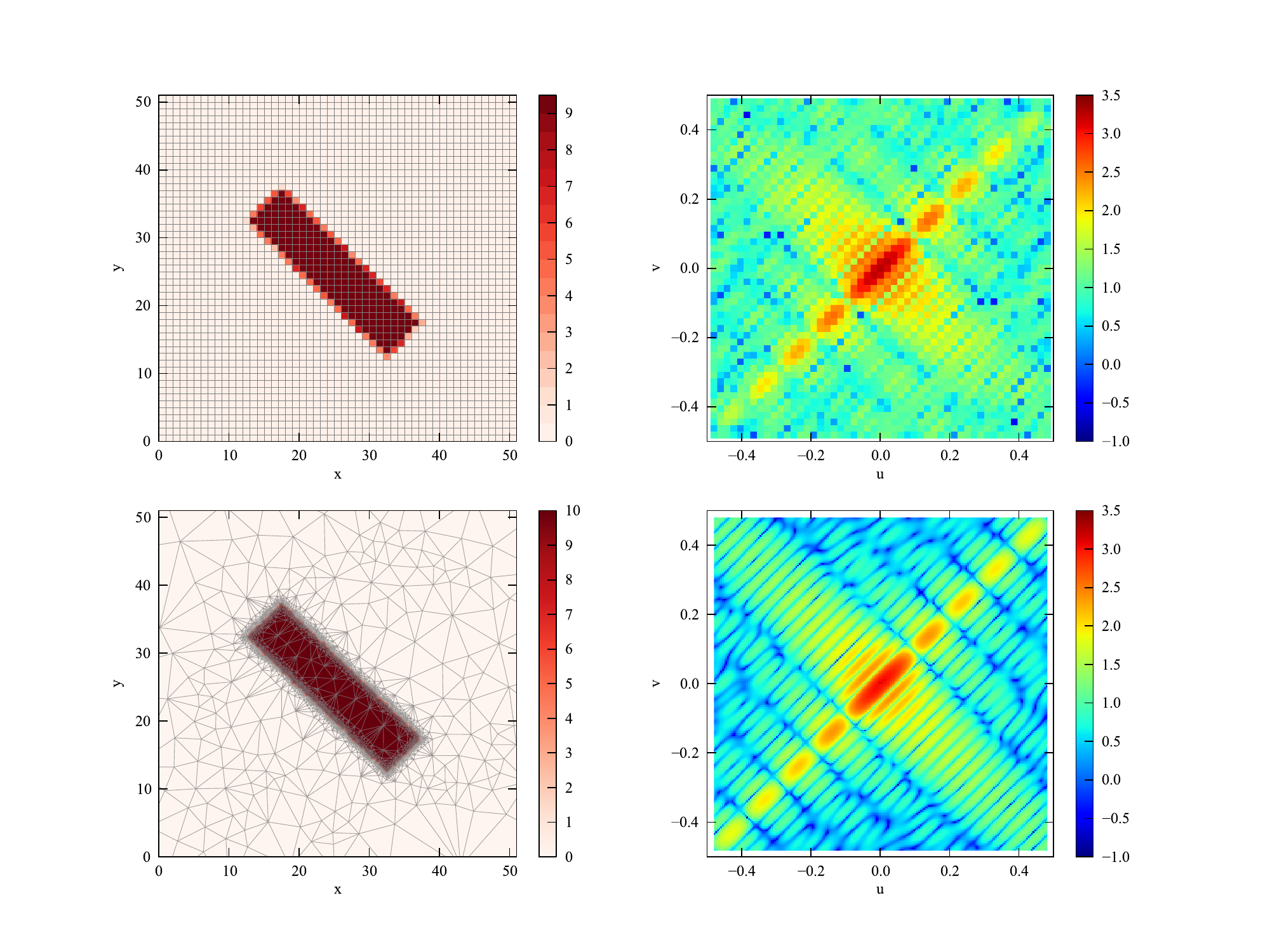}}
\caption{Similar to Fig.~\ref{transforms1}, but for a rectangular shape 
   function. In this example we have used anti-aliasing on the raster image in 
   order to smooth out the sharp edges and minimize the artefacts due to the 
   staircase effect of the square pixels.  }\label{transforms2}
\end{figure*}

First we show the Fourier transform of two simple shapes, a circular annulus 
(Fig.~\ref{transforms1}) and an inclined rectangle (Fig.~\ref{transforms2}). 
Both shapes have been imaged on a 50$\times $50 (=2500) pixel raster map as well 
a on a 2500 point trixel map. The resulting images are shown in the left column 
of Fig.~\ref{transforms1} and \ref{transforms2}. The points for the trixel 
images have been chosen so that most of the points fall inside the shape and 
just across the edge. In fact, for the rectangle, only the edge has been mapped 
with a high density of points, whereas the constant inside and outside of the 
rectangle has a relatively low point density. The result is a very clearly 
defined shape for both the annulus and the rectangle as compared to the raster 
version where the edge appears rough (or indeed, pixelated). We have furthermore 
used anti-aliasing with 16 rays per pixel in the raster version of the rectangle 
in order to smooth out the edges. This means that the pixels along the edges are 
coloured according to the fraction of rays that are randomly sampled within each 
pixel that falls inside the rectangle. This anti-aliasing requires an additional 
800 rays. When deciding where to place the rays for the triangulated images, it 
is generally not possible to do emission weighted ray-tracing, since the regions 
of strong emission are not know a priori. However, if the source structure is 
known, at least to a certain degree (which is the case for the rectangle and 
annulus cases shown here), one can place rays accordingly. When using the LIME 
code, the positions of the grid points are already weighted by the source 
structure and the projection of the grid points can then be used as a proxy for 
the distribution of the rays. Many ray-tracing codes already make use of some 
sort of ``clever'' ray 
placement~\citep[e.g.,][]{Dullemond2000,2009ApJ...704.1482P} from which a 
triangulated image could be formed rather than remapping the rays onto a raster 
image.

The right column of Fig.~\ref{transforms1} and \ref{transforms2} shows the 
corresponding Fourier transforms, FFT in the case of the raster images and the 
method derived above in the case of the trixel images. It is clear that for both 
cases, the two different Fourier transformation methods produce basically the 
same result, but because our method is not resolution limited, the resulting 
Fourier transform comes out in a much, much higher resolution (actually higher 
than what can be displayed in a raster contour plot).  The FFT of the annulus 
shows clear asymmetric edge effects which are completely absent in trixel 
Fourier transform. The anti-aliasing which is used to smooth out the edges in 
the rectangle example of Fig.~\ref{transforms2} does not improve the FFT 
significantly.

\begin{figure*}
\centerline{\includegraphics[width=18cm]{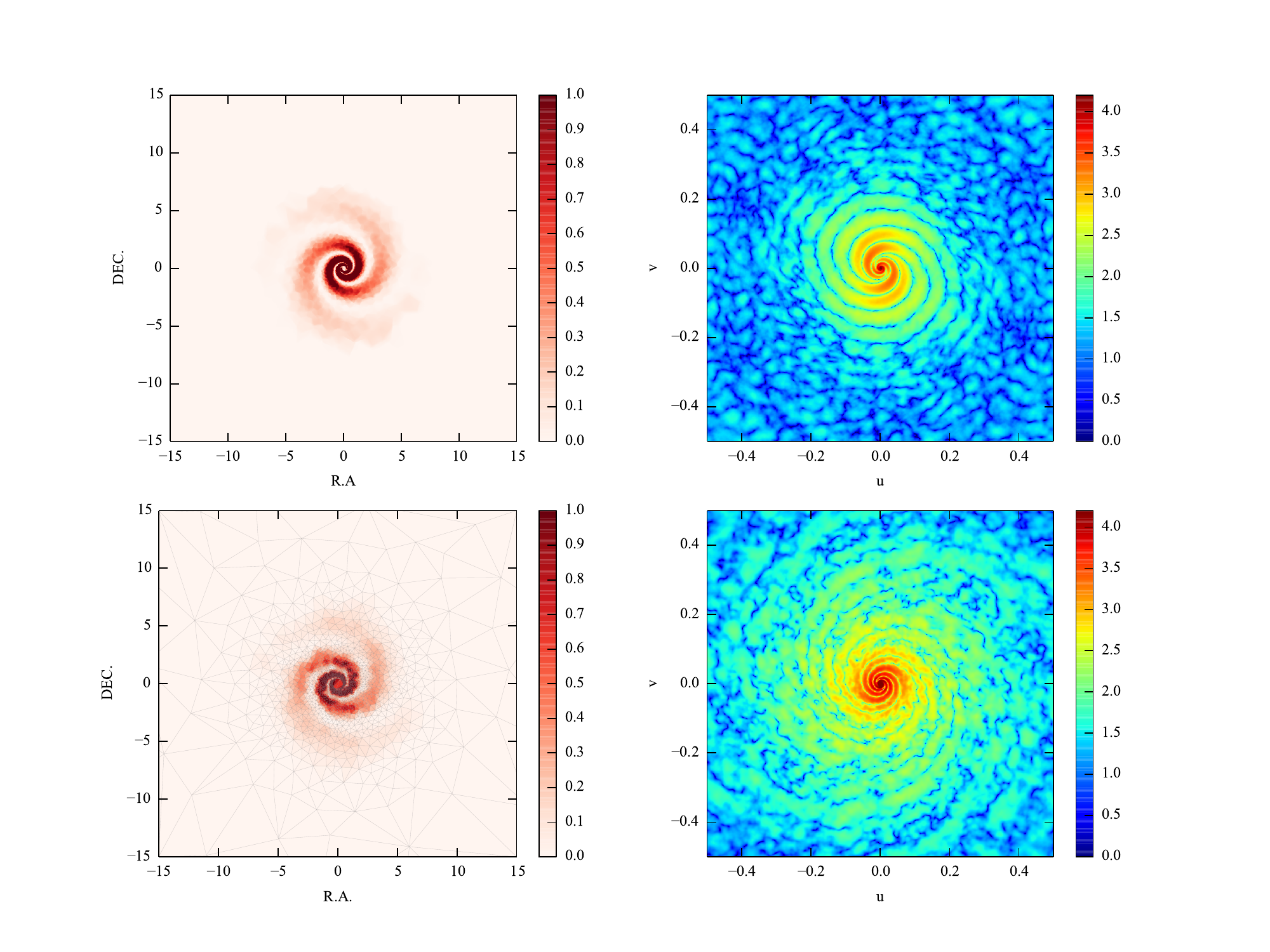}}
\caption{Similar to Fig.~\ref{transforms1} but for a radiative transfer model of 
   a toy disk model.}
\label{twhya}
\end{figure*}

The next example is a more realistic case, an astronomical pseudo-model, based 
on a real protoplanetary disk model, with the addition of an arbitrary spiral 
density perturbation. The spiral was chosen in order to 
introduce spatial variations on all scales and is in no way assumed to describe 
any physical reality, hence pseudo-model. However, the underlying disk model is 
consistent with current models of the disk around the T Tauri star TW Hydra (as 
described in, e.g., \citet{2012ApJ...744..162A}) in terms of mass, mass 
distribution, size, distance, temperature, etc. The radiation transfer model has 
been calculated by the LIME code and the solution has been ray-traced to produce 
an image of the continuum at 1 mm. Figure~\ref{twhya} shows the resulting raster 
image and trixel image. The resolution and image size of the raster image has 
been chosen to minimize the artifacts in the visibilities which can be measured 
by ALMA, given a source distance of 50 parsec, assuming a synthesized beam size
of 0.1 arcsec and a maximum recoverable scale of 15 arcsec. The resulting raster
image consistst of $10^3 \times 10^3$ pixels. The trixel image is made out of as
few as possible rays, and hence trixles, which still produces a Fourier transform
which is as good as the FFT. In this case the number of rays are 5000 or a factor
of 200 fewer rays than what is needed for the raster image. If we use fewer rays 
for the trixel image the Fourier transform no longer compares well to the FFT. This,
however, is not due to a limitation of the method, but simply because we no longer
describe the source structure well enough. An important thing to notice is that,
despite the fewer rays used in the trixel image, we can trace the spirals in the Fourier
transform to much greater uv-distances, although the spiral do tend to become noisier.

Comparing the Fourier transforms, we see that there is quite some difference 
between the FFT and our method. A test shows that if raster images are made in 
progressively higher resolution, the FFT approaches the Fourier transform 
produced by our method, and so the morphologic difference the two Fourier 
transforms shown here are only due to the inadequate resolution in the raster 
image.

Finally, we have made a small resolution study of our method and compared this 
with the performance of the FFT. For this, we used a 2D unit disk which has the 
well-known analytic Fourier transform,

\begin{eqnarray}
F(u,v) = \frac{j_1(2 \pi \sqrt{u^2+v^2})}{\sqrt{u^2+v^2}}.
\end{eqnarray}

The unit disk has been mapped onto a regular pixel grid, using anti-aliasing, 
that is, with a fractional shading of pixels proportionally to the fraction of 
the pixel area that is covered by the disk. For the trixel version, the unit 
disk has been mapped similarly to the example in Fig.~\ref{transforms1}.

\begin{figure*}
\includegraphics[width=8cm]{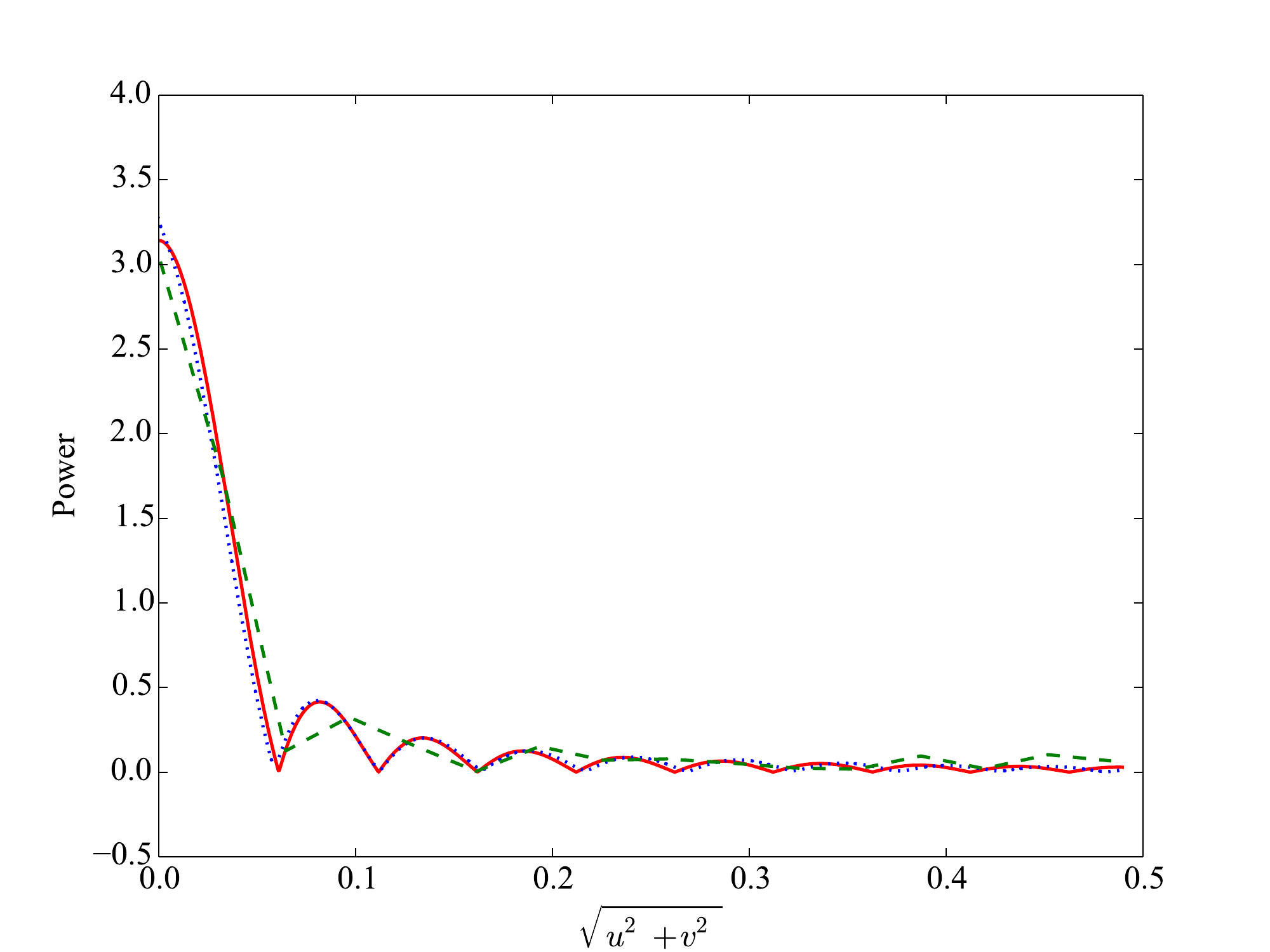}
\includegraphics[width=8cm]{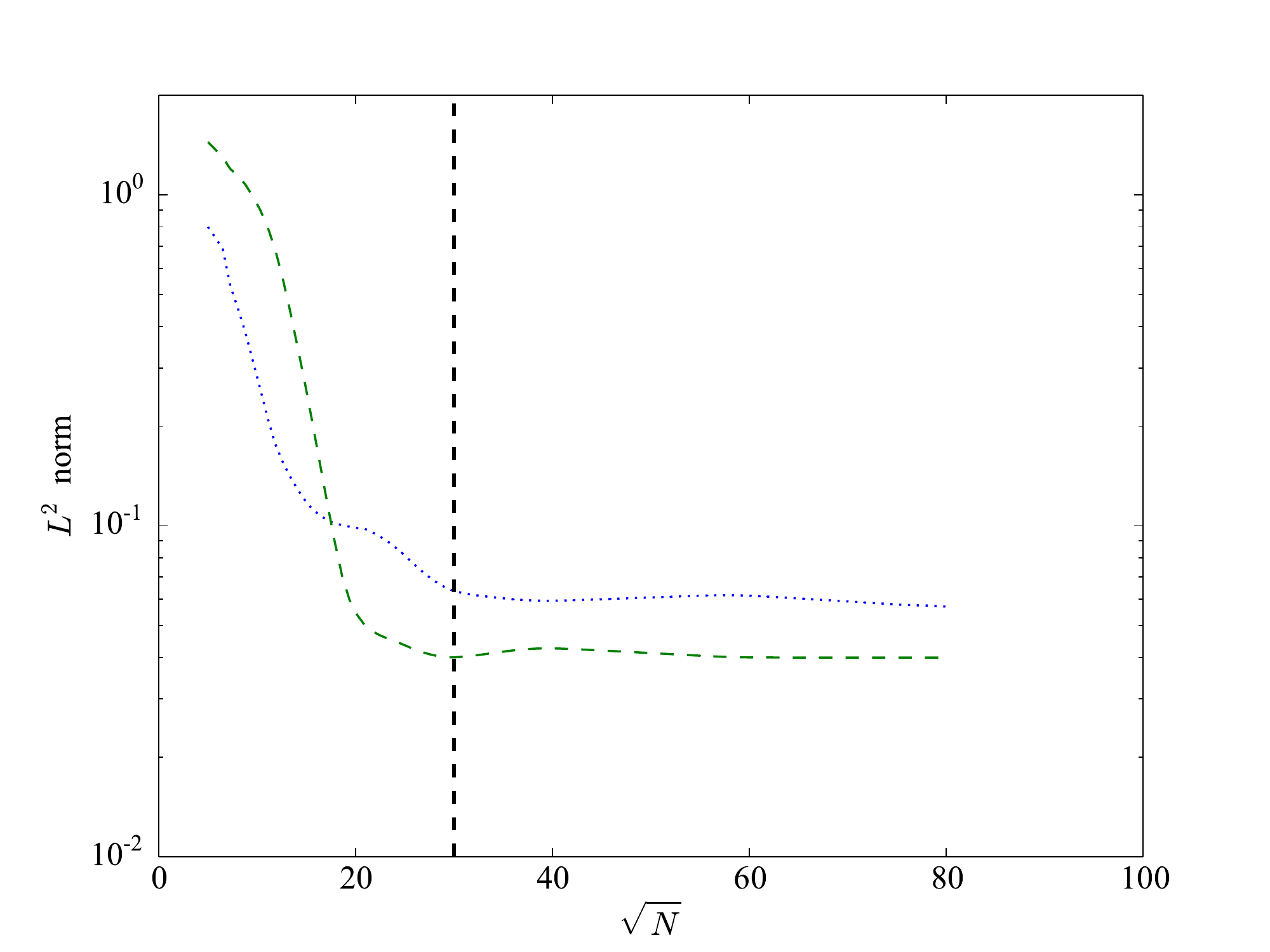}
\caption{The left panel shows a comparison between the anaytic Fourier transform 
   of a unit disk (full red line) and the results of FFT (dashed green) and the trixel method (dotted blue). 
   The right panel shows the L$^2$ norm of both methods as function of resolution.
   $N$ refers to the number of pixels in the case of FFT and the number of rays
   in the case of the trixel method.}
   \label{airy}
\end{figure*}

The result of the resolution study is shown in Fig~\ref{airy}, where the full red 
curve shows the analytic solution and the dotted blue and dashed green curves show the trixel 
method and FFT respectively. The left panel shows the solutions at a resolution 
of $30^2$ pixels for the FFT version and $30^2$ rays for the trixel method. The 
right panel in Fig.~\ref{airy} shows the L$^2$ norm (the Euclidean distance between solutions) of both methods as a function of 
increasing resolution. Both methods seem to have converged at around $N=30$. 
After convergence, the L$^2$ norm of FFT is slightly better than the trixel method, 
but judged by the fit in the main panel, both methods give very accurate 
results. The phase is slightly offset for larger uv-spacings in the trixel 
method, but on the other hand, the FFT over-produces the power by a small amount 
on the smallest spatial scales and obviously, the FFT method has a significantly 
lower resolution in uv-space than what can be achieved by the method presented 
here.

\section{Conclusions}
In this paper we have presented a method to create radiative transfer model 
images in arbitrary resolution and very high dynamic range using a finite, and 
much smaller number of rays than is needed for a raster image in comparable 
resolution. The method uses an unstructured (possibly random) distribution of 
rays out of which a Delaunay triangulation is calculated. Each Delaunay triangle 
is easily Fourier transformed using Eq.~\ref{mcinturff1}.

Unfortunately, Eq.~\ref{mcinturff1} becomes very time consuming for large or 
``complete'' sets of uv-spacings. The Fourier transformation method presented 
here requires O(N$^2$) operations which from a performance point of view is 
vastly outperformed, particularly for large N, by FFT which requires O(N log N) 
operations. However, like Eq.~\ref{mcinturff1}, ray-tracing in a 3-D radiative 
transfer model is also an O(N$^2$) process in the number of pixels per axis and 
so what is gained in speed from using FFT is quickly lost again from the 
increased ray-tracing time in order to reach high enough image resolution. One 
could consider taking an unstructured, and therefore high-resolution, set of 
rays and remap it onto a very high resolution raster in order to perform an FFT. 
Such a remapping, however, can potentially also be quite time consuming on its own
and it still requires a somewhat arbitrary choice of minimum and maximum scale 
to be made. Increasing the number of pixels dramatically 
has the further disadvantage of producing very large FITS files, in particular 
when doing spectral line images, where the spectral axis can potentially hold 
hundreds of channels. For the example in Fig~\ref{twhya}, the FFT on the raster 
image is done in less than a few seconds (which means that the computation time 
is dominated by I/O and other overhead), whereas the Fourier transform on the 
trixel image takes a total of 1.5 minutes. However, this Fourier transform has about four 
times higher resolution than the FFT. In order to reach the same resolution in 
uv-space with FFT, the image has to be ray-traced at four times the resolution. 
The FFT operation on the higher resolution raster image does not take noticeable
longer, but the ray-tracing time, in this example, goes from about 20 seconds to 
about 4 minutes and this does not include the additional time requirement when 
adding anti-aliasing in order to improve the image quality. It is also possible 
to lower the computation time significantly for the unstructured Fourier 
transform method, when comparing a model to interferometric data, by only 
calculating the uv-points which corresponds to the observed uv-spacings, 
rather than calculating complete sets of uv-points. Equation~\ref{sum_tri} is also
trivially parallelisable which helps to speed up the calculations since most modern 
computers have multiple cores.

There is currently no image container for unstructured triangulated images 
although the FITS format could in principle be used. One option would be to 
build the Fourier transformer directly into the ray-tracing code and let the 
code output a uv-FITS file rather than having the Fourier transformation be a 
post-processing tool that works on outputted images. Alternatively, trixels can 
be stored in standard FITS format as tabulated data.

\bibliographystyle{mn2e.bst}
\bibliography{references}
\end{document}